\documentclass[12pt]{iopart}

\usepackage{iopams}  
\usepackage{url}
\usepackage{bm}
\usepackage{amssymb}
\usepackage{graphicx}
\usepackage{float}
\usepackage{floatflt}
\usepackage{slashed}
\begin{document}

\title[Saving supersymmetry and dark matter WIMPs]{Saving supersymmetry and
dark matter WIMPs -- \\
a new kind of dark matter candidate with well-defined mass and couplings}

\author{Roland E. Allen}
\address{Department of Physics and Astronomy, Texas A\&M University \\
College Station, Texas 77843, USA}

\begin{abstract}
Since neither supersymmetry nor dark matter WIMPs have yet been observed,
pessimism about their reality has been growing. Here we discuss a new
supersymmetric theory and a new dark matter candidate which are naturally
consistent with current experimental results, but which imply a plethora
of new phenomena awaiting discovery within the foreseeable future.
\end{abstract}

%\maketitle

Although there is so far no evidence for either supersymmetry (susy)~\cite
{Baer-Tata,Kane,no-susy} or dark matter particles~\cite
{susy-DM-1996,Silk,Strigari,Rauch,Freese,Roszkowski,Aprile}, both these
extensions of Standard Model physics remain well-motivated~\cite
{Baer-Barger1,Nath,Peskin}. In the words of Ref.~\cite{Nath}, ``We note in passing
that the unification of gauge coupling constants is satisfied to a good
degree of accuracy in models with scalar masses lying in the tens of TeV
...'' However, there is increasing tension between experiment and the
proposal that susy can explain dark matter~\cite
{Olive1,Olive2,Baer-Barger2,Baer-Barger3}. In the words of Ref.~\cite
{Baer-Barger3}, ``Supersymmetric models of particle physics have been under
assault from both collider search experiments and direct and indirect dark
matter detection experiments.'' Even for more general weakly interacting
massive particles (WIMPs), it was felt eight years ago that~\cite{Bertone}
``With the advent of the Large Hadron Collider at CERN, and a new generation
of astroparticle experiments, the moment of truth has come for WIMPs: either
we will discover them in the next five to ten years, or we will witness
their inevitable decline.''

Recently we introduced a new kind of dark matter candidate~\cite{DM1} which
inevitably follows from a fundamental theory~\cite{statistical}, but for
which we will here simply postulate the phenomenological model given below
in Eqs. (\ref{S1S2}) and (\ref{S1}). In some ways the particle proposed here
resembles the lowest-mass neutralino of susy, the linear combination of
neutral fermionic superpartners that is currently the most popular of
specific dark matter candidates: It is charge neutral, with only weak gauge interactions; and it has a spin of 1/2 and an R-parity of -1,
making it stable if its mass is less than that of the lowest mass
superpartner. (The R-parity is $\left( -1\right) ^{3B+L+2s}$, with both the
present particle and the neutralino having spin $s=1/2$, baryon number $B=0$
, and lepton number $L=0$. If these three quantities are additively
conserved, the R-parity $R_{P}$ is multiplicatively conserved.) 

There are also major differences, however, that will enable experiment to
distinguish the present particle from a neutralino (as well as other
candidates): As discussed near the end of this paper, its gauge interactions --
i.e., couplings to $W$ and $Z$ bosons -- are in a sense weaker than those of
the neutralino, since they are either second-order or momentum-dependent.
I.e., each coupling is proportional to $V^{\mu }V^{\prime}_{\mu }$, 
$P_{\mu }V^{\mu }$, or $V^{\mu } \, p_{\mu }$, where $V_{\mu }$ or $V^{\prime}_{\mu }$ 
represents a gauge boson field $W_{\mu }^{\pm }$ or $Z_{\mu }$, 
$P_{\mu }$ is the momentum for this field
(or a $W^{\pm }$ or $Z$ particle), and $p_{\mu }$ is the momentum of the
dark matter field or particle proposed here. The couplings of a neutralino,
on the other hand, are first-order, proportional to just $V_{\mu }$. The
relatively weak gauge interactions of the present particles may then explain why
dark matter particles have not yet been detected: In both direct detection
(thorugh collisions with nuclei) and indirect detection (through
annihilation in space) they are WIMPs with relatively low momenta $p_{\mu }$; 
and in collider-based detection, they will be relatively hard to create.
Nevertheless, as described immediately below, they constitute an ideal dark matter
candidate in other respects.

Both supersymmetry and the new particles proposed here are inevitable
consequences of the fundamental theory of Ref.~\cite{statistical}, and the
theory cannot even be formulated without these features,. It is therefore gratifying
that the lowest-mass 
of these particles automatically turns out to have many desirable features for a dark
matter candidate: As a WIMP with a mass at or near the electroweak scale
(since it is comparable to that of the recently discovered Higgs boson), it
should have been produced in the early universe with about the right
abundance to explain the astronomical observations. With an R-parity of $-1$, 
it will be stable in the later universe, provided that its mass lies below
that of the lowest energy superpartner (also with R-parity $=-1$). Through
its coupling to $W$ and $Z$ bosons, it can in principle be observed within
the foreseeable future in collider, direct detection, and indirect detection
experiments. It also appears to be the only dark matter candidate with a
well-defined mass plus well-defined couplings. Specifically, its mass $m_{H}$
is limited by $m_{H}\leq m_{h}$, where $m_{h} = 125$~GeV/$c^2$ is the mass of the
Higgs boson, with $m_{H}=m_{h}$ in the very simplest case. This inequality
places it within the optimal range for direct detection by many experiments 
that are currently very active.

A key feature of the present theory is the implication that all Higgs-like
fields have a richer structure than in standard physics: Each scalar Higgs
boson is interpreted as an amplitude mode of a 4-component field, roughly
analogous to the Higgs/amplitude mode observed in superconductors. As
discussed in Refs.~\cite{DM1} and \cite{statistical}, the usual picture of a
scalar Higgs condensate and scalar Higgs bosons is regained at the energies
that have so far been experimentally explored, but the new spin 1/2
particles proposed here should be observable at energies above the threshold
for their creation in pairs.

As mentioned above, the fundamental theory of Ref.~\cite{statistical}
unambiguously predicts supersymmetry, which retains its role in (i) protecting
the mass of the Higgs boson from a divergence imposed by radiative
corrections and (ii) unifying the coupling constants of the nongravitational
forces at high energy. The theory then predicts that susy particles will
eventually be seen at sufficently high energies.

The theory also predicts, however, a new sector of spin 1/2 particles. If
the lightest of these particles $H^{i}\ $has a lower mass than the lightest
superpartner, it will be the most stable of particles with an R-parity
of -1, undercutting any potential susy dark matter candidate. This is a very
plausible scenario because the candidate proposed here has a mass of  $\leq
125$~GeV/$c^2$, whereas the masses of susy candidates can range up to 1~TeV/$c^2$ or
higher.

The action for the new fields and particles proposed here\ is 
\begin{eqnarray}
S_{\Phi }=S_{1}+S_{2} 
\label{S1S2}
\end{eqnarray}
where 
\begin{eqnarray}
S_{1}=\int d^{4}x\,\left( \frac{1}{2}\left( i\overline{\sigma }^{\mu }D_{\mu
}\Phi \left( x\right) \right) ^{\dag }\,\left( i\sigma ^{\nu }D_{\nu }\Phi
\left( x\right) \right) +h.c.\right) \;.
\label{S1}
\end{eqnarray}
The notation is defined in Refs.~\cite{DM1} and \cite{statistical}: $S_{2}$
consists of mass terms (which may arise from many quadratic and quartic
terms even in relatively simple supersymmetric Higgs models), $h.c.$ means
Hermitian conjugate, the $\sigma $ matrices have their usual definitions
(and are always implicitly multiplied by an appropriate identity matrix),
and $D_{\mu }$ is the usual covariant derivative for the electroweak gauge
fields. After symmetry breaking it has the form~\cite{peskin}
\begin{eqnarray}
\hspace{-0.5in}
D_{\mu }=\partial _{\mu }-i\frac{g}{\sqrt{2}}\left( W_{\mu }^{+}T^{+}+W_{\mu
}^{-}T^{-}\right) -i\frac{g}{\cos \theta _{w}}Z_{\mu }\left( T^{3}-\sin
^{2}\theta _{w}\,Q\right) -ieA_{\mu }\,Q\;.
\label{der}
\end{eqnarray}
This exhibits the coupling to the $W^{\pm}$ and $Z$ fields (and to the photon 
field $A_{\mu}$ for charged particles), with the standard notation defined in Ref.~\cite{peskin}.

If $\Phi $ is written in the form 
\begin{eqnarray}
\Phi =\left( 
\begin{array}{c}
\Phi ^{R} \\ 
\Phi ^{R^{\prime} }
\label{8-component}
\end{array}
\right) 
\end{eqnarray}
where the fields $\Phi ^{R }$ and $\Phi ^{R^{\prime} }$ are defined below, it is convenient to use the
same Weyl representation as for Dirac fields, where 
\begin{eqnarray}
\gamma ^{\mu }=\left( 
\begin{array}{cc}
0 & \sigma ^{\mu } \\ 
\overline{\sigma }^{\mu } & 0
\end{array}
\right) \;,
\end{eqnarray}
so that (after integration by parts with neglect of boundary terms) 
\begin{eqnarray}
\hspace{-1in}
S_{1} &=&\int d^{4}x\,\frac{1}{2}\left( \Phi ^{R \, \dag }\left(
x\right) i\sigma ^{\mu }D_{\mu }\,i\overline{\sigma }^{\nu }D_{\nu }\Phi
^{R }\left( x\right) +\Phi ^{R^{\prime} \, \dag }\left( x\right) i
\overline{\sigma }^{\mu }D_{\mu }\,i\sigma ^{\nu }D_{\nu }\Phi ^{R^{\prime} }
\left( x\right) \right) +h.c.  \label{eq7.52x} \\
\hspace{-0.7in} &=&\int d^{4}x\,\left( -\frac{1}{2}\Phi ^{\dag }\left( x\right) \gamma ^{\mu
}D_{\mu }\,\gamma ^{\nu }D_{\nu }\Phi \left( x\right) \right) +h.c. \\
\hspace{-0.7in} &=&-\int d^{4}x\,\frac{1}{2}\Phi ^{\dag }\left( x\right) \,\slashed{D}
^{2}\,\Phi \left( x\right) +h.c.  \label{eq14.5}
\end{eqnarray}

According to a result~\cite{schwartz} that can easily be extended to the
nonabelian case~\cite{peskin}, we have 
\begin{eqnarray}
\slashed{D}^{2}=-D^{\mu }D_{\mu }+S^{\mu \nu }F_{\mu \nu }
\label{identity}
\end{eqnarray}
with a $(-+++)$ convention for the metric tensor. The second term gives an
addition to standard physics, involving the total field strength tensor $
F_{\mu \nu }$ for the electroweak gauge fields and the Lorentz generators $
S^{\mu \nu }$ which act on Dirac spinors: 
\begin{eqnarray}
\mathcal{S}_{1}=\int d^{4}x\,\left( \frac{1}{2}\Phi ^{\dag }\left( x\right)
D^{\mu }D_{\mu }\Phi \left( x\right) -\frac{1}{2}\Phi ^{\dag }\left(
x\right) \,S^{\mu \nu }F_{\mu \nu }\,\Phi \left( x\right) \right) +h.c.
\label{full-S1} 
\end{eqnarray}
where 
\begin{eqnarray}
S^{\mu \nu }=\frac{1}{2}\sigma ^{\mu \nu }\;
\end{eqnarray}
or \cite{schwartz} 
\begin{eqnarray}
S^{kk^{\prime }}=\frac{1}{2}\varepsilon _{kk^{\prime }k^{\prime \prime
}}\left( 
\begin{array}{cc}
\sigma ^{k^{\prime \prime }} & 0 \\ 
0 & \sigma ^{k^{\prime \prime }}
\end{array}
\right) \quad ,\quad S^{0k}=-\frac{i}{2}\left( 
\begin{array}{cc}
\sigma ^{k} & 0 \\ 
0 & -\sigma ^{k}
\end{array}
\right) \;\;.
\end{eqnarray}
This can be rewritten in terms of the ``magnetic'' and ``electric'' fields $B_{k}$ and $E_{k}$ defined by 
\begin{eqnarray}
F_{kk^{\prime }}=-\varepsilon _{kk^{\prime }k^{\prime \prime }}B_{k^{\prime
\prime }}\quad ,\quad F_{0k}=E_{k}
\end{eqnarray}
as~\cite{DM1,statistical} 
\begin{eqnarray}
S_{1}=\int d^{4}x\,\left( \Phi ^{\dag }\left( x\right) D^{\mu }D_{\mu }\Phi
\left( x\right) +\Phi ^{\dag }\left( x\right) \,\overrightarrow{B}\cdot 
\overrightarrow{\sigma }\,\Phi \left( x\right) \right) \;.
\label{full-S1-B}
\end{eqnarray}
As discussed in \cite{DM1}, the new features of this term will have
observable effects only at high energy (or in extremely small radiative
corrections at lower energy), and in conjunction with the new spin 1/2
particles predicted here. 

The form (\ref{full-S1}), involving $F_{\mu \nu }$, might be regarded as more fundamental than (\ref{full-S1-B}), 
since the fields $\Phi ^{R}$ and $\Phi^{R^{\prime }}$ are coupled through the mass matrix considered below.

In the treatment above, the summation convention has been used, but in the
remainder of this paper summations will always be explicitly indicated and 
not implied over repeated indices.

In Ref.~\cite{statistical}, the forms (\ref{S1}) and (\ref{full-S1}) are
derived for all Higgs-like boson fields (including those at a GUT scale),
and each component $\Phi ^{r}$ of $\Phi $ consists itself of two complex
components. Here we will
consider the mimimal case of two Higgs doublets, both with weak hypercharge $Y=1$. 
Then in (\ref{8-component}), $\Phi ^{R}$ and $\Phi^{R^{\prime }}$ are both doublets, with four complex components each.

Let us begin with a general treatment of the amplitude (scalar) modes for either the neutral 
(weak isospin $T_3=-1/2$) or positively charged ($T_3=+1/2$) part of the doublets.

We then have two fields $\Phi ^{r}$ and $\Phi^{r^{\prime }}$ with the same gauge quantum 
numbers ($Y=1$,  $T_3=-1/2$ or $+1/2$), each with 2 components in the present 
description. They can be grouped together in a 4-component object 
\begin{eqnarray}
\Phi ^{r,r^{\prime }}=\left( 
\begin{array}{c}
\Phi ^{r} \\ 
\Phi ^{r^{\prime }}
\end{array}
\right) \;.\;
\end{eqnarray}
(This is either the neutral or the charged part of the 8-component $\Phi$ in (\ref{8-component}).)

We can achieve a scalar condensate (for a neutral field) and scalar excitations (for neutral or charged fields) by requiring that 
\begin{eqnarray}
\Phi ^{r,r^{\prime }}=\phi ^{r,r^{\prime }}\left( 
\begin{array}{c}
\chi ^{r} \\ 
\chi ^{r^{\prime }}
\end{array}
\right) 
\end{eqnarray}
where $\phi ^{r,r^{\prime }}$ is a complex scalar and the 2-component
spinors satisfy 
\begin{eqnarray}
\chi ^{r^{\prime }\,\dag }\overrightarrow{\sigma }\,\chi ^{r^{\prime
}}=-\chi ^{r\,\dag }\overrightarrow{\sigma }\,\chi ^{r}
\label{scalar}
\end{eqnarray}
or 
\begin{eqnarray}
\Phi ^{r,r^{\prime }\,\dag }\overrightarrow{\sigma }\,\Phi ^{r,r^{\prime
}}=0\;.
\end{eqnarray}
The components of the 3-vector $\overrightarrow{\sigma }$ are the Pauli
matrices (and, again, it is understood that $\overrightarrow{\sigma }$ is
multiplied by an appropriate identity matrix). It follows that 
\begin{eqnarray}
\Phi ^{r,r^{\prime }\,\dag }\left( x\right) \,\overrightarrow{B}\cdot 
\overrightarrow{\sigma }\,\Phi ^{r,r^{\prime }}=0
\end{eqnarray}
and with the normalization 
\begin{eqnarray}
\chi ^{r\,^{\prime }\dag }\chi ^{r^{\prime }}=\chi ^{r\,\dag }\chi ^{r}=1/2
\end{eqnarray}
the contribution to (\ref{full-S1}) is just 
\begin{eqnarray}
\mathcal{S}_{1}^{r,r^{\prime }}=\int d^{4}x\,\phi ^{r,r^{\prime \, *}
}\left( x\right) D^{\mu }D_{\mu }\,\phi ^{r,r^{\prime }}\left( x\right) \;.
\end{eqnarray}
In general, then, the constraint (\ref{scalar}) results in the standard action for a
scalar boson field. 

Standard physics is thus regained if the internal degrees of
freedom in $\Phi $ are not excited. As will be seen below, and was
emphasized previously~\cite{DM1,statistical}, this requires the
production of a pair of massive spin 1/2 particles, though processes that are 
either second-order or momentum dependent.

Now let us consider in more detail the neutral fields and their condensate. The two $Y=1$ 
fields will be labeled 1 and 2, with a specific spin configuration labeled by 
$\uparrow $ or $\downarrow $. The condensate has the form
\begin{eqnarray}
\Phi_0 =\phi_0 \left( 
\begin{array}{c}
\chi _{\uparrow }^{1} \\ 
\chi _{\downarrow }^{2}
\end{array}
\right)                                                                             
\\ 
\chi _{\downarrow }^{2\,\dag }
\overrightarrow{\sigma }\,\chi_{\downarrow }^{2} 
=-\chi _{\uparrow }^{1\,\dag }\overrightarrow{
\sigma }\,\chi _{\uparrow }^{1}\quad ,\quad \chi _{\downarrow
}^{2\,\dag }\chi _{\downarrow }^{2}=\chi _{\uparrow }^{1\,\dag }\chi
_{\uparrow }^{1}=1/2 \; .
\label{constraint}
\end{eqnarray}
There are two independent amplitude-mode excitations, 
which are respectively aligned and anti-aligned with the condensate:
\begin{eqnarray}
\Delta \Phi=\Delta \phi \left( 
\begin{array}{c}
\chi _{\uparrow }^{1} \\ 
\chi _{\downarrow }^{2}
\end{array}
\right) \quad ,\quad \Delta \overline{\Phi }=\Delta \overline{\phi } \left( 
\begin{array}{c}
\chi _{\downarrow }^{1} \\ 
\chi _{\uparrow }^{2}
\end{array}
\right) \;.
\label{phi}
\end{eqnarray}

More generally, there are separate spin 1/2 excitations of the 1 and 2
fields, with four independent possibilities since each 2-component field has two spin
degrees of freedom: 
\begin{eqnarray}
\Delta \widetilde{\Phi }^{1\uparrow } &=&\left( 
\begin{array}{c}
\widetilde{H}^{1\uparrow } \\ 
0
\end{array}
\right) \quad ,\quad \Delta \widetilde{\Phi }^{1\downarrow }=\left( 
\begin{array}{c}
\widetilde{H}^{1\downarrow } \\ 
0
\end{array}
\right)  \label{states 1}
\\
\Delta \widetilde{\Phi }^{2\uparrow } &=&\left( 
\begin{array}{c}
0 \\ 
\widetilde{H}^{2\uparrow }
\end{array}
\right) \quad ,\quad \Delta \widetilde{\Phi }^{2\downarrow }=\left( 
\begin{array}{c}
0 \\ 
\widetilde{H}^{2\downarrow }
\end{array}
\right) \;. 
\label{states 2}
\end{eqnarray}
To avoid complexity of notation, we will use $\widetilde{H}^{i}$ to represent the 
4-component fields shown above.

$\Phi $, like the standard Higgs field, has self-interactions. With
the constant term and higher-order terms neglected, and for the neutral
fields alone, the self-interaction Lagrangian density has the form
\begin{equation}
-\mathcal{L}_{2}^{0}=\Delta \widetilde{\Phi }^{\dag }\widehat{M^{2}}\,\Delta \widetilde{
\Phi }\;.
\end{equation}
Here $\widehat{M^{2}}$ is a constant $4 \times 4$ Hermitian matrix which
is roughly analogous to the mass matrix for quark or lepton fields. 
If $\Delta \widetilde{\Phi} $ is written as a sum of fields $H^{i}$ that are eigenstates 
of $\widehat{M^{2}} $ -- see (\ref{mass1}) and (\ref{mass2}) below -- with the amplitudes 
required to represent $\Delta \widetilde{\Phi} $, we obtain the diagonal form
\begin{equation}
-\mathcal{L}_{2}^{0}=\sum\limits_{i}M_{i}^{2}\,H^{i\,\dag }H^{i}\;.
\end{equation}
A mass eigenstate $H^{i}$ is a 
linear combination of the ``flavor'' and spin states of (\ref{states 1}) 
and (\ref{states 2}), just as a neutrino mass eigenstate is a linear combination 
of $\nu_e$, $\nu_{\mu}$, and $\nu_{\tau}$ flavor states.

At each fixed $x$, the four 
orthogonal $4$-component eigenvectors $H^{i}\left( x\right) $ satisfy 
\begin{eqnarray}
\widehat{M^{2}}\,H^{i}\left( x\right)  = M_{i}^{2}H^{i}\left( x\right)  
\label{mass1}
\\
H^{i^{\prime }\,\dag }\left( x\right) H^{i}\left( x\right)  = H^{i\dag
}\left( x\right) H^{i}\left( x\right) \delta ^{i^{\prime }i}\;.
\label{mass2}
\end{eqnarray}
It will be assumed that $\widehat{M^{2}}$ also permits amplitude-mode
eigenstates $\Phi ^{j}$ with eigenvalues $m_{j}^{2}$:
\begin{eqnarray}
\widehat{M^{2}}\,\Phi ^{j}\left( x\right) =m_{j}^{2}\Phi ^{j}\left( x\right) \; .
\end{eqnarray}
The $H^{i}\left( x\right) $ are a complete set of eigenvectors, so 
\begin{eqnarray}
\;\Phi ^{j\,}\left( x\right) =\sum\limits_{i}\,c_{ji}\,H^{i}\left( x\right) 
\end{eqnarray}
and
\begin{eqnarray}
\Phi ^{j\,\dag }\left( x\right) \,\widehat{M^{2}}\,\Phi ^{j\,}\left(
x\right) =\sum\limits_{i^{\prime }\,i}c_{ji^{\prime }}^{\ast
}\,c_{ji}\,H^{i\prime \,\dag }\left( x\right) \,\widehat{M^{2}}\,H^{i}\left(
x\right) 
\end{eqnarray}
or
\begin{eqnarray}
m_{j}^{2}\,\Phi ^{j\,\dag }\left( x\right) \,\Phi ^{j\,}\left( x\right)
=\sum\limits_{i}M_{i}^{2}\,c_{ji}^{\ast }\,c_{ji}\,H^{i\,\dag }\left(
x\right) \,H^{i}\left( x\right) \;.
\end{eqnarray}
Replacing $\widehat{M^{2}}$ by the identity matrix gives
\begin{eqnarray}
\Phi ^{j\,\dag }\left( x\right) \,\Phi ^{j\,}\left( x\right)
=\sum\limits_{i}\,c_{ji}^{\ast }\,c_{ji\,}H^{i\,\dag }\left( x\right)
\,H^{i}\left( x\right) \;.
\end{eqnarray}

The above equations hold at fixed $x$. We can, however, take the
fields $H^{i}\left( x\right) $ to be functions which have the same
dependence on $x$ as the given amplitude mode $\,\Phi ^{j\,}\left( x\right) $
or its scalar field $\,\phi ^{j\,}\left( x\right) $ (and which 
therefore do not satisfy the equation of motion for a physical
particle with mass $M_{i}$). Then the coefficients $c_{ji\,}$ are constant,
and if we take $\Phi ^{j\,}$ and the $H^{i}$ to be normalized
single-particle excitations,
\begin{eqnarray}
\hspace{-1.1in}
m_{j}^{2}\,=m_{j}^{2}\,\int d^{3}x\,\Phi ^{j\,\dag }\left( x\right) \,\Phi
^{j\,}\left( x\right) =\sum\limits_{i}M_{i}^{2}\,c_{ji}^{\ast }\,c_{ji}\int
d^{3}x\,\,H^{i\,\dag }\left( x\right) \,H^{i}\left( x\right)
=\sum\limits_{i}M_{i}^{2}\,c_{ji}^{\ast }\,c_{ji}
\end{eqnarray}
and 
\begin{eqnarray}
\hspace{-0.7in}
1=\int d^{3}x\,\Phi ^{j\,\dag }\left( x\right) \,\Phi
^{j\,}\left( x\right) =\sum\limits_{i}c_{ji}^{\ast }\,c_{ji}\int
d^{3}x\,\,H^{i\,\dag }\left( x\right) \,H^{i}\left( x\right)
=\sum\limits_{i}c_{ji}^{\ast }\,c_{ji} \; .
\end{eqnarray}
I.e., the mass squared of a scalar Higgs boson $m_{j}^{2}$ is equal to the
average mass squared for all the spin $1/2$ degrees of freedom associated with it:
\begin{eqnarray}
m_{j}^{2}=\left\langle M_{i}^{2}\right\rangle _{j}\equiv \frac{
\sum\limits_{i}M_{i}^{2}c_{ji}^{\ast }\,c_{ji}}{\sum
\limits_{i}c_{ji}^{\ast }\,c_{ji}} \; .
\end{eqnarray}

The mass $M_{H}$ of the dark matter particle proposed here is the lowest of
the $M_{i}$, so the above equation predicts 
\begin{eqnarray}
M_{H}\leq m_{h}  \label{mass}
\end{eqnarray}
where $m_{h}=125$~GeV/$c^2$ is the lowest mass of a Higgs boson (i.e., the mass of
the only Higgs boson so far observed). If the scalar and spinor mass
eigenvalues turned out to be matched, one could have $M_{H}=m_{h}$, but the
above inequality is the more robust prediction.

The above treatment is actually valid for either the neutral or charged components of two complex
Higgs doublets, so there are 4 neutral and 4 charged scalar degrees of
freedom. There are then 3
would-be Goldstone bosons, one charged Higgs, and 3 neutral Higgses. 
In the simplest interpretation, $\phi $ and $\overline{\phi } $ of (\ref{phi}) 
are respectively the condensed and uncondensed fields of the Higgs basis~\cite{Silva}.

In addition to all these Higgses, we now have 4 neutral and 4 charged spin 1/2
particles. Let us now consider the gauge interactions and processes that can lead
to discovery in direct detection, indirect detection, and collider
experiments. For the spin 1/2 fields in the present two-doublet model, (\ref{full-S1}) has the form
\begin{eqnarray}
\hspace{-0.5in}
S_{H}=\sum\limits_{i}\int d^{4}x\,\left( H^{\,\dag }\left( x\right) D^{\mu
}D_{\mu }H\left( x\right) -\left( \frac{1}{2}H^{\dag }\left( x\right) \,S^{\mu \nu
}F_{\mu \nu }\,H\left( x\right) +h.c. \right) \right) 
\end{eqnarray}
where $H$ is the 8-component field consisting of both neutral and charged spin $1/2$ excitations. 

There is a complication in the terms that couple the charged fields  
to the neutral fields, because their mass-squared operators 
$\widehat{M_0^{2}}$ and $\widehat{M_+^{2}}$  are in general different. 
The original ``flavor'' and spin states are then separately expanded in 
mass eigenstates $H^{+i}$ and $H^{0i}$, causing the mass eigenstates 
to be mixed in their interactions with the $W^{\pm}$.
This is analogous to the complication that leads to the CKM matrix for
quarks. Here, however, we are primarily  concerned with only the 
charge-neutral particles and fields.
 
Since $D_{\mu}$ has the form (\ref{der}) 
and the field strength has the form 
\begin{eqnarray}
F_{\mu \nu }^{a}=\partial _{\mu }V_{\nu }^{a}-\partial _{\nu }V_{\mu
}^{b}+g_{V}\,f^{abc}V_{\mu }^{b}V_{\nu }^{c}
\end{eqnarray}
one can read off the most relevant gauge interaction terms as $H^{i\,\dag }W^{+\mu }W_{\mu
}^{-}H^{i},$ $H^{i\,\dag }Z^{\mu }Z_{\mu }H^{i},$ $H^{i\,\dag }P^{\mu
}Z_{\mu }H^{i},$ $H^{i\,\dag }Z^{\mu }p_{\mu }H^{i}$, with $-i\partial _{\mu
}\longrightarrow P_{\mu }$ or $p_{\mu }$. One can then construct the Feynman
diagrams for the various processes that are relevant to experiment, and in
principle calculate cross-sections. In the following, we will use the generic name $H$ 
for a spin $1/2$ field or particle of the kind considered here (with $h$ reserved 
for the corresponding scalar fields and particles).

Direct detection: The elastic scattering of an $H$ particle (or antiparticle 
$\overline{H}$)  is fundamentally mediated by virtual $Z^{0}$
or $W^{+},W^{-}$ exchange with quarks. The details of either coherent or
incoherent scattering off a complete nucleus are, of course, more complicated.

Indirect detection:  $H$, $\overline{H}$ annihilation in space can produce a virtual 
$Z^{0}$,  $W^{+},W^{-}$ pair, or $Z^{0},Z^{0}$ pair, which then decay through Standard Model
processes to the particles that can be detected by, e.g., Fermi-LAT or
AMS-02. The cleanest signature would be  $Z^{0}\longrightarrow f \, \bar{f} \longrightarrow 2\gamma $ 
with each gamma-ray photon $\gamma $ having the energy of the dark matter
particle $H$. The various other signatures have been extensively anticipated
and explored, but for different annihilation cross-sections and branching ratios.

Collider detection: In the LHC, quark collisions can produce, e.g., $H$, 
$\overline{H}$ pairs (in a variety of processes involving $Z^{0}$ or $W^{\pm }
$ exchange) which will show up as missing transverse energy. (There
are, of course, no production mechanisms analogous to the production of
Higgs bosons through gluon fusion mediated by the top quark, because there
are no possible direct couplings of these spin 1/2 particles to a pair of
fermions.)

In summary, with well-defined weak-interaction couplings, 
an R-parity of $-1$ (providing stability), and a mass that is $\leq 125$~GeV/$c^2$, 
the particles predicted
here are in many respects ideal dark matter candidates. However, their gauge 
couplings are in a sense weaker than those expected for the most popular of
the previous candidates, and this fact may explain why dark matter particles
have so far eluded detection. The theory that predicts these new particles -- which are 
associated with an extended version of the Higgs sector -- also unambiguously
predicts supersymmetry. The fact that susy has also not yet been observed is
then attributed to a higher energy scale for superpartners than has been
explored so far. Perhaps most important, the present theory predicts a
plethora of new neutral and charged particles, and new physics, to be
discovered at collider energies that could be available in the foreseeable
future.

\end{document}